\documentclass[a4paper]{article}

\usepackage{icrc2013}
\usepackage[english]{babel}

\def\vyp#1#2#3{\textbf{#1} (#2) #3}

\title{Testing models of new physics with UHE air shower observations}

\shorttitle{Models of new physics}

\authors{
Jeffrey D. Allen$^{1}$ and Glennys R. Farrar$^{1}$
}

\afiliations{
$^1$ Center for Cosmology and Particle Physics, Department of Physics, New York University, NY, NY 10003, USA\\
}

\email{jda292@nyu.edu, gf25@nyu.edu}

\abstract{
Several air shower observatories have established that the number of muons produced in UHE air showers is significantly larger than that predicted by models. We argue that the only solution to this muon deficit, compatible with the observed $X_{max}$ distributions, is to reduce the transfer of energy from the hadronic shower into the EM shower, by reducing the production or decay of $\pi^0$s. We present four different models of new physics, each with a theoretical rationale, which can accomplish this. One has a pure proton composition and three have mixed composition. Two entail new particle physics and suppress $\pi^0$ production or decay above LHC energies. The other two are less radical but nonetheless require significant modifications to existing hadron production models -- in one the changes are only above LHC energies and in the other the changes extend to much lower energies.  We show that the models have distinctively different predictions for the \emph{correlation} between the number of muons at ground and $X_{max}$ in hybrid events, so that with future hybrid data it should be possible to discriminate between models of new physics and disentangle the particle physics from composition.

}


\begin{document}
\maketitle

\section{Introduction}
Measurements of the density of muons at ground in ultra-high energy (UHE) air-showers performed by hybrid observatories, first at HiRes-MIA~\cite{casamia} and more recently at the Pierre Auger Observatory~\cite{prevNmu, prevHoriz,curNmu,curHorizontal}, have revealed that there is a significant deficit of muons in Monte Carlo (MC) simulations of air showers. The number of muons in the data is greater than that predicted using even iron initiated air-showers. Explaining this is made more challenging by the measurements of the distribution of the depth of shower maximum, $X_{max}$. Measurements performed at the Pierre Auger Observatory (PAO) and  Telescope Array (TA) both show that, depending on the hadronic model used to interpret the $X_{max}$ data, the mean mass at 10 EeV is light to intermediate~\cite{augerXmax, augerInterXmax,taXmax, hiresXmax}.

In the present study, we investigate potential resolutions to this discrepancy. We begin by exploring how generic properties of hadronic interactions are constrained by independent measurements of the density of muons at 1000 m from the shower core, $N_{\mu}$, and $X_{max}$. We then present four schematic models of hadronic interactions which represent different methods to simultaneously fit measurements of both the mean $X_{max}$ and $N_{\mu}$. Each model can be tuned to reproduce the observed $X_{max}$ distribution and mean $N_{\mu}$. Fortunately, air shower observatories can differentiate the four models by studying the correlation between the $X_{max}$ and the $N_{\mu}$ of hybrid air showers. Measurements of this nature should be feasible at both the PAO and TA, especially with upgrades to the muon sensitivity.

\section{Constraints on Hadronic Interactions}
$N_{\mu}$ and $X_{max}$ are sensitive to several properties of hadronic interactions. Some properties, such as the primary cosmic ray mass composition and multiplicity of secondary particles, impact both these observables, while other properties impact only one or the other. By studying how MC predictions for $N_{\mu}$ and $X_{max}$ behave under modifications to various hadronic interaction properties, we identify potential modifications which could resolve the muon deficit.

The mean $N_{\mu}$ is primarily sensitive to the multiplicity, the $\pi^0$ energy fraction (the fraction of incident energy carried by $\pi^0$s in hadronic interactions), and the primary mass. The mean $X_{max}$ is primarily sensitive to the cross-section, elasticity, multiplicity, and primary mass. This dependence appears in the hadronic extension of the Heitler model~\cite{hadHeitler}, and has been studied quantitatively~\cite{ulrich, eposppbar}. Table~\ref{tabHadIntMods} summarizes the qualitative impact that changing each property has on $N_{\mu}$ and $X_{max}$.

\begin{table}[t]
\begin{center}
\caption{A summary of the dependence of $N_{\mu}$ and $X_{max}$ as various properties of the hadronic interactions are increased, with all others held fixed. }
\vspace{0.25cm}
\begin{tabular}{|l|c|c|}
\hline Property Increased & Change in $N_{\mu}$ & Change in $X_{max}$ \\ \hline
Cross-section   & --  & Decreased \\ \hline
Elasticity   & --  & Increased \\ \hline
Multiplicity & Increased  & Decreased \\ \hline
Primary Mass & Increased  & Decreased \\ \hline
$\pi^0$ Eng. Frac. & Decreased  & -- \\ \hline
\end{tabular}
\label{tabHadIntMods}
\end{center}
\vspace{-0.27in}
\end{table}

To explore how changes in these properties affect the mean $N_{\mu}$ and $X_{max}$, we modify the secondary particles of the hadronic interaction model in the MC simulations. Modifications are made in a similar manner to that in~\cite{ulrich}. The simulations are performed using the SENECA~\cite{seneca} air shower simulation with EPOS 1.99~\cite{EPOS} as the underlying hadronic event generator (HEG), although any other HEG could be used as the starting point. The primary energy in all simulations in this paper is $10^{19}$ eV; except as noted modifications are performed at energies above $10^{17}$ eV and become stronger with increasing energy.

\begin{figure}[t]
  \centering
  \includegraphics[width=0.46\textwidth]{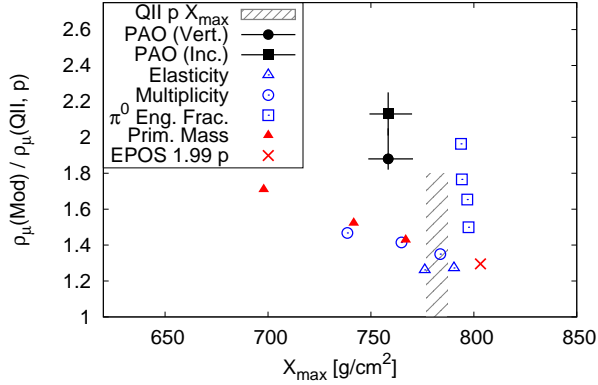}
  \caption{The mean $X_{max}$, and density of muons at 1000 m relative to that predicted by QGSJET-II-03 for proton showers, for various compositions and modifications to hadronic interactions as detailed in the text. 
The PAO datapoints for the mean $X_{max}$ and muon density at $10^{19}$ eV are from \cite{augerXmax} and \cite{prevNmu,prevHoriz}; the grey-hatching indicates $< X_{max}> $ for QGSJET-II-03 protons, which is compatible with HIRES data~\cite{hiresXmax}. 
}
  \label{figMuXMaxHadIntMods}
  \vspace{-0.15in}
 \end{figure}

Fig.~\ref{figMuXMaxHadIntMods} shows how the mean $X_{max}$ and $N_{\mu}$ change under a suite of modifications, starting from EPOS 1.99 for protons shown with an ``x''.\\
$\bullet$ Composition (solid triangles): He, C, Fe.\\
$\bullet$ Multiplicity (open circles):  Non-leading secondary particles are split into multiple particles, conserving energy; the probability of splitting is tuned to produce a 100\% to 700\% increase in the multiplicity.\\
$\bullet$ Elasticity (open triangles): Interactions which have a leading particle with $x_{F}> 0.4$ have a chance of being re-simulated with a probability tuned to reduce the number of elastic events by 50\% and 80\%.\\
$\bullet$ $\pi^0$ energy fraction (open squares): Forward pions are converted into baryons and other pions are converted into kaons, with a common probability varied from 8\% to 60\%.

As expected, increasing the multiplicity or primary mass decreases the $X_{max}$ and increases $N_{\mu}$.  However these mechanisms can play at most a partial role in solving the muon deficit, because $X_{max}$ becomes too shallow before $N_{\mu}$ is sufficiently increased.  The best hope for resolving the muon deficit is in decreasing the $\pi^0$ energy fraction, because that is the only modification which increases the mean $N_{\mu}$ without encountering any restrictions from the $X_{max}$ observations \footnote{Increasing the lateral spread of the muons by increasing the $<p_{t}>$ can increase the muon density at 1000m for some particular zenith angle, but results in a zenith-angle dependence of the ground signal that is incompatible with observations (JDA and GRF, in preparation).}.

\section{Description of New Models}
We have developed four schematic models which rely primarily upon modifying the fraction of energy which is transfered to decaying $\pi^0$s in order to fit both the mean $N_{\mu}$ and $X_{max}$ observed at the Auger Observatory. The models are implemented by modifying the secondary particles of EPOS 1.99. The energies of the initial interactions of an ultra-high energy air-shower are well above those achieved at accelerators. The center of mass energy of a 10 EeV proton incident upon a nucleon in the atmosphere is 137 TeV, and many secondary interactions are above the energy of the LHC. This justifies taking considerable freedom in exploring potential, new physics scenarios. We investigate new physics scenarios to uncover signatures of new physics and to explore a broad range of mechanisms which have the potential to solve the muon deficit.

 \begin{figure}[t]
  \centering
  \includegraphics[width=0.46\textwidth]{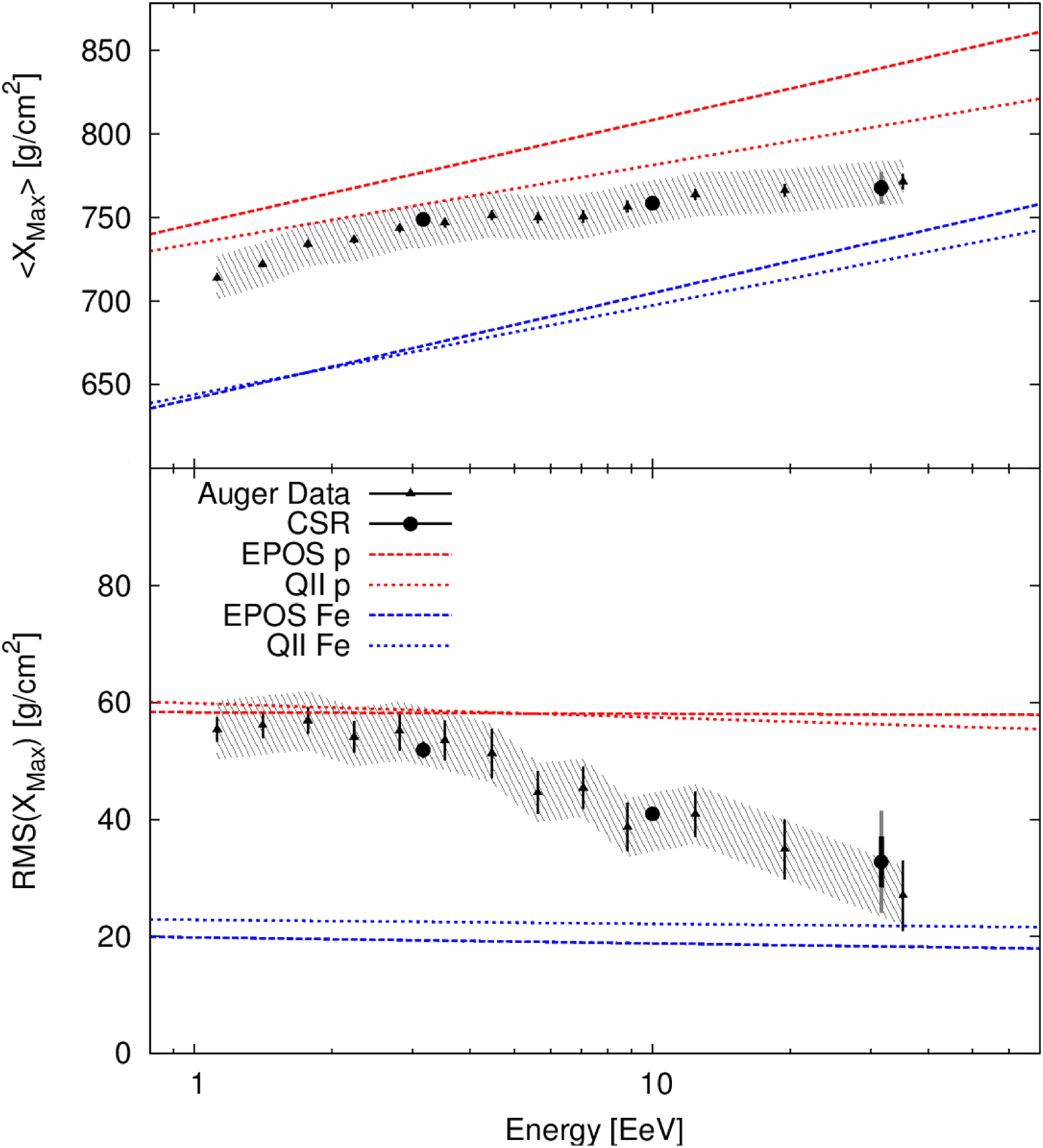}
  \caption{The energy dependence of the mean and RMS of $X_{max}$ in the CSR model compared to Auger data~\cite{csr}.}
   \label{figCSR-ER-RE}
    \vspace{-0.1in}
 \end{figure}

\subsection{Chiral Symmetry Restoration}
In the Chiral Symmetry Restoration (CSR) model~\cite{csr}, we imagine that at the energy densities achieved in some hadronic interactions in UHE air showers, chiral symmetry is restored and the production of pions becomes greatly suppressed. In the CSR model, the primary cosmic rays are protons in order to achieve the highest energy density. The energy density of an interaction is determined by the impact parameter. We use the elasticity of the interaction as a tracer of the impact parameter; when the elasticity of the interaction is below a certain threshold, the interaction enters the CSR phase.

More modifications are necessary to realize an acceptable pure-proton scenario than simply a reduction of the production of pions; the average $X_{max}$ predicted by EPOS using proton primaries is deeper than observations by both the Auger Observatory and TA.  In the CSR phase, the multiplicity is increased and elasticity decreased. The cross section is rapidly increased at high energies to reduce the $RMS$ of the predicted $X_{max}$ distribution. The tunable parameters of the CSR model include the strength of the pion production suppression, the elasticity threshold for entering the CSR phase, the elasticity of CSR interactions, and the increase in the proton-Air cross-section. Through suitable adjustment of the energy dependence of the parameters, an acceptable mean $X_{max}$ and $RMS(X_{max})$ can be found for all energies (Fig. \ref{figCSR-ER-RE}); see ~\cite{csr} for details. The CSR model thus provides an example of a proton-only model which can fit air shower observables~\cite{csr}.

\subsection{Pion decay suppression}
Pion decays in air shower MCs are treated as if they take place in a vacuum. However, in the rest frame of high energy pions, the atmosphere is in fact a very dense medium. We postulate that pion decay could be suppressed through interactions with the dense air medium~\cite{pidecsup}. In the pion decay suppression (PDS) model, pion decay is suppressed at high energies. The impact this has on air shower development is similar in effect to simply decreasing pion production: since the $\pi^0$s do not decay, they do not feed the electromagnetic shower. There is only one tunable parameter of the model, which is the energy above which pion decay is suppressed for a reference air density. The primary mass composition is assumed to be mixed in order to provide freedom to fit the $X_{max}$ distribution.

 \begin{figure}[t]
  \centering
  \includegraphics[width=0.46\textwidth]{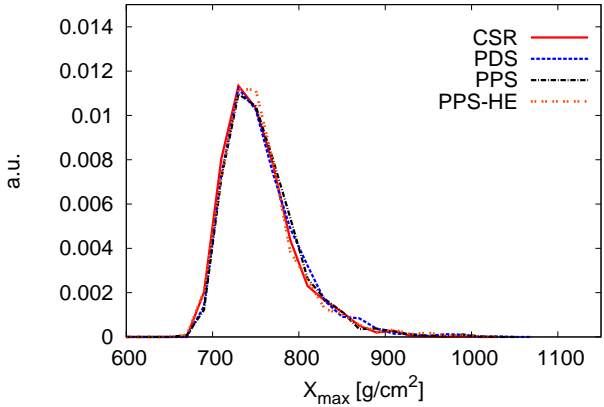}
  \caption{The distribution of $X_{max}$ for the four models.}
  \label{figXMaxDists}
    \vspace{-0.15in}
 \end{figure}

 \begin{figure}[t]
  \centering
  \includegraphics[width=0.46\textwidth]{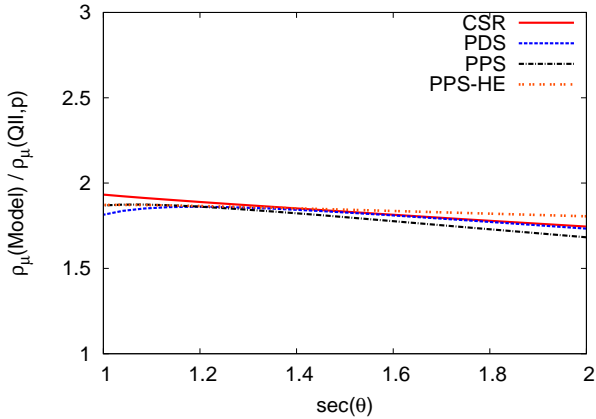}
  \caption{The density of muons at 1000 m, relative to QGSJET-II-03 proton showers, as a function of zenith angle for the four models tuned to remove the muon deficit at $38^{\circ}$; the model-to-model variation in zenith angle dependence is within systematic uncertainties. }
  \label{figNMuSecT}
    \vspace{-0.1in}
 \end{figure}

\subsection{Pion production suppression}
The popular HEGs, such as QGSJET-II~\cite{QGSJET}, SIBYLL~\cite{SIBYLL} and EPOS, predict a wide range of $\pi^0$ energy fractions. For example, QGSJET-II-03 predicts that ~25\% of all the energy in secondary particles are carried by $\pi^0$s and $\eta$s, while for EPOS 1.99 and SIBYLL 2.1 this is closer to 20\%. This difference in the models persists at all energies. It is thus possible that the fraction of energy carried by pions in hadronic interactions may be less than any of the models currently predict.

We consider two variants of a pion production suppression model: i) (PPS) at all energies, pions in the forward direction, chosen to be above a pseudorapidity of 5 at LHC energy, are converted to baryons, and all pions, regardless of pseudorapidity, are converted to kaons, with a common probability that is a tunable parameter, and ii) (PPS-HE) the same modification is made but only for interactions with incident energy above $10^{17}$ eV. The high-energy variant is introduced to explore the impact of performing the pion production suppression in different energy regimes of shower development and would be similar to modifications to the string percolation probabilities~\cite{stringPercolation}, or if heavy flavor production is enhanced in kinematic regimes where quark masses may be insignificant. As in the pion decay suppression model, the primary mass composition is assumed to be mixed.

\subsection{Comparison between models and data}
Figs.~\ref{figXMaxDists} and \ref{figNMuSecT} show that each of these models can be tuned through their various parameters to fit both the $X_{max}$ data and the magnitude and zenith angle dependence of the density of muons at 1000 m, at $10^{19}$ eV.  The mean $N_{\mu}$ constrains primarily the strength of the pion production/decay suppression in the models. The mean $X_{max}$ constrains the mass composition in the PDS, PPS, and PPS-HE models and the cross section and elasticity in the CSR model. There is sufficient freedom in the models to tune them to produce nearly identical predictions for $X_{max}$ and $N_{\mu}$.  The differences in zenith angle dependence of the muon density at 1000m are too small to be used to discriminate between the models, given systematic uncertainties and the fact that it is the total signal including the EM component which is presently measured.  

\section{\texorpdfstring{The $N_{\mu}$ - $X_{max}$ Plane}{The Nmu - Xmax Plane}}
As demonstrated in Sec. 3, we can construct models which fit the muon ground density and $X_{max}$ data using various deviations from standard HEGs. Fortunately, the four models can be discriminated by observing correlations between the depth of shower maximum and the number of muons at ground for an ensemble of showers.

To discriminate between the models, the distribution of showers in the $N_{\mu}$-$X_{max}$ plane must be observed. Any observable which is related to the total number of muons in the showers is suitable, generically called $N_{\mu}$ here; for definiteness we continue to use the density of muons at 1000 m. The correlation of $N_{\mu}$ and $X_{max}$ is primarily sensitive to two basic properties of hadronic interactions: the mass composition, and the energy threshold for the suppression of pion decay and production.

The number of muons produced in air showers is determined, in part, by the number of generations between the energy at which pion production begins and the energy at which pions decay, since in each generation energy is lost to the electromagnetic sub-shower~\cite{ehpShowerReview,hadHeitler}. In general, iron and other heavy primaries have fewer generations and thus produce more muons.  This causes a large negative correlation between $N_{\mu}$ and $X_{max}$ when the composition is mixed, and a weak correlation in the case of a proton-only composition~\cite{younkrisse}. However, when pion production is suppressed above an energy threshold, this tends to equalize the number of generations between initial pion production and decay, and thus produce a similar number of muons. This causes the strength of the correlation to decrease as the degree of the pion production/decay suppression is increased.

\begin{figure}[t]
  \centering
  \includegraphics[width=0.46\textwidth]{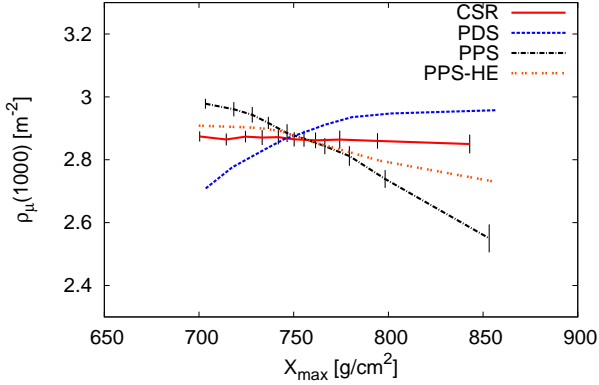}
  \caption{The density of muons at 1000 m as a function of $X_{max}$ for the four template models. The error bars show the variance in samples of 800 hybrid events, which is a number achievable at the PAO and TA.}
  \label{figSX4Models}
        \vspace{-0.1in}
 \end{figure}

The four models span a wide range of primary compositions and mechanisms for suppressing pion production or decay and, thus, have different correlation strengths between $N_{\mu}$ and $X_{max}$.  Fig.~\ref{figSX4Models} shows the average $N_{\mu}$ as a function of $X_{max}$ and its variance in ensembles of 800 simulated events, for 10 EeV showers at a zenith angle of $38^{\rm o}$, for each of the four schematic models.  These simulations were done at a fixed zenith angle and energy, but the same statistical power could be achieved by normalizing showers to a fiducial energy and angle.   The negative correlation between $N_{\mu}$ and $X_{max}$ is strongest when the modification is made at all energies, and weaker when the modification is applied only at high energy. The PPS and PPS-HE models show a negative correlation while the CSR model shows almost no correlation. Finally, the PDS model actually shows a positive correlation: showers with a shallow $X_{max}$ produce fewer muons than showers with a deep $X_{max}$.   Thus the models can be discriminated at high significance with presently realizable datasets.

\begin{figure}[t]
  \centering
  \includegraphics[width=0.46\textwidth]{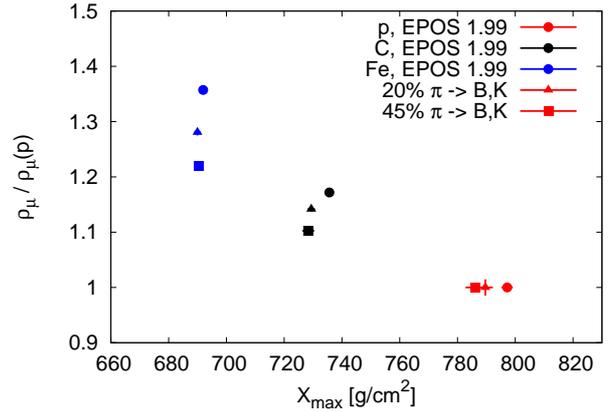}
  \caption{The dependence of the density of muons at 1000 m, relative to protons, on pion production at high energy.}
  \label{figSX-Pi0HiEng}
        \vspace{-0.1in}
 \end{figure}

The behavior of the PDS and PPS models demonstrates that the correlation is sensitive to the energy threshold of the pion modifications. This is made explicit in Fig. \ref{figSX-Pi0HiEng}, which compares the average $N_{\mu}$ and $X_{max}$ of iron, carbon, and proton initiated showers, for different degrees of pion production suppression at high energy. As pion production suppression is increased, the relative difference between the mean $N_{\mu}$ in iron and proton showers decreases.

\section{Conclusion}
We argue that the muon deficit in simulations of air showers indicates that the hadronic models are incorrectly predicting the fraction of energy which is transfered to the electromagnetic sub shower. Changing the $\pi^0$ energy fraction or suppressing pion decay are the only modifications which can be used to increase the number of muons at ground without coming into conflict with the $X_{max}$ observations.

We have developed four schematic models of hadronic interactions, all of which are capable of correctly describing the $X_{max}$ distributions and number of muons at ground. They utilize both pure proton and mixed primary compositions, which demonstrates the need for models to correctly describe all air shower observables before they are used to interpret the primary mass composition.

The four models can be distinguished by observations of the correlation between the number of muons $N_{\mu}$ and $X_{max}$, for an ensemble of hybrid events. This correlation provides a crucial new observable for determining the nature of UHE air showers.  Existing hybrid datasets may already be large enough to rule out some explanations of the muon excess.

\noindent {\bf Acknowledgements:}  
The authors are members of the Pierre Auger Collaboration
and acknowledge with gratitude innumerable valuable discussions with colleagues. 
This research was supported by NSF-PHY-1212538.  

\end{document}